\documentclass{jfm}

\usepackage{graphicx}
\usepackage{epstopdf}
\usepackage{natbib}
\usepackage{amssymb}
\usepackage{amsbsy}

\def\Re{{\rm Re}}

\def\Ca{{\rm Ca}}
\def\Fr{{\rm Fr}}
\def\St{{\rm St}}

\newcommand{\e}{\mathrm{e}}
\newcommand{\dx}{\mathrm{d}}
\newcommand{\JB}{\mathrm{J_0}}

\title[Initial surface deformations during impact on a liquid pool]{Initial surface deformations during impact on a liquid pool}
\author[W. Bouwhuis, M.H.W. Hendrix, D. van der Meer, and J.H. Snoeijer]{Wilco Bouwhuis$^1$, Maurice H.W. Hendrix$^{1,2}$, Devaraj van der Meer$^1$, and Jacco H. Snoeijer$^{1,3}$}
\affiliation{$^1$ Physics of Fluids Group, Faculty of Science and Technology, University of Twente, 7500 AE Enschede, The Netherlands,\\$^2$ Laboratory for Aero and Hydrodynamics, Delft University of Technology, Leeghwaterstraat 21, NL-2628 CA Delft, The Netherlands,\\$^3$ Mesoscopic Transport Phenomena, Eindhoven University of Technology, Den Dolech 2, 5612 AZ Eindhoven, The Netherlands}

\pubyear{2014}
\volume{?}
\pagerange{??}
\begin{document}

\maketitle

\begin{abstract}
A tiny air bubble can be entrapped at the bottom of a solid sphere that impacts onto a liquid pool. The bubble forms due to the deformation of the liquid surface by a local pressure buildup inside the surrounding gas, as also observed during the impact of a liquid drop on a solid wall. Here we perform a perturbation analysis to quantitatively predict the initial deformations of the free surface of the liquid pool as it is approached by a solid sphere. We study the natural limits where the gas can be treated as a viscous fluid (Stokes flow) or as an inviscid fluid (potential flow). For both cases we derive the spatio-temporal evolution of the pool surface, and recover some of the recently proposed scaling laws for bubble entrapment. When inserting typical experimental values for the impact parameters, we find that the bubble volume is mainly determined by the effect of gas viscosity.
\end{abstract}

\section{Introduction}\label{sec:Introduction}

\begin{figure}
\centering
\includegraphics[width=0.99\textwidth]{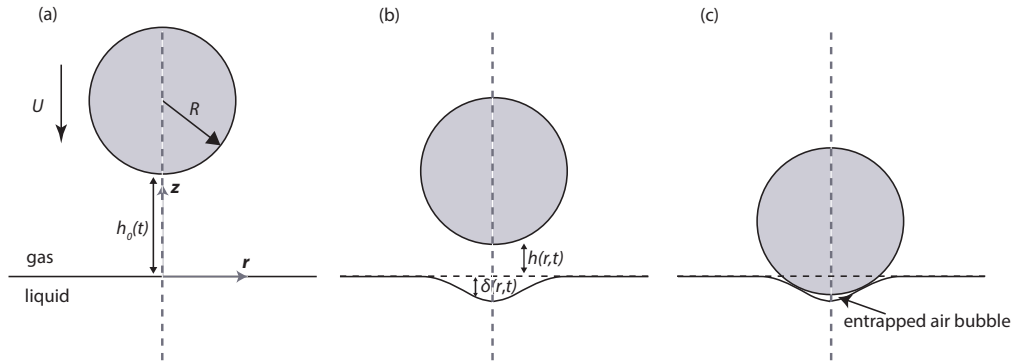}
\caption{(a) A solid sphere (radius $R$) approaches a liquid surface with velocity $U$. The gap height between the bottom of the sphere and the undisturbed water level ($z=0$) is $h(r,t)$, $r$ and $t$ being the radial coordinate and time, respectively, with $h(0,t)=h_0(t)$. (b) While the sphere moves downwards, the pool deflects by a small amount $\delta(r,t)$, as a result of the local pressure build-up in the air that is squeezed out. In the limit where $\delta\ll h_0$, which typically is valid up to very close to the impact time, the profiles are computed analytically. (c) This mechanism will finally result in air bubble entrapment.}
\label{Setting}
\end{figure}

The phenomena resulting from solid-body impacts on liquid surfaces are widely studied because of their omnipresence in nature and industry \citep{Korobkin1988, Howison1991, Korobkin2008, Do-Quang2009, Deng2009, Marston2011, Hicks2012, Moore2014}. These involve splashing, jet formation, cavity formation, and air bubble entrapment. The mechanism for entrapment of tiny, micrometer-sized air bubbles between the solid object and the pool is due to a mechanism similar to that of the impact of a liquid drop on a solid surface \citep{Smith2003, Dam2004, Thoroddsen2005, Driscoll2011, Mandre2012, Bouwhuis2012, Klaseboer2014} or of a drop onto a liquid pool \citep{Yiantsios1990, Hicks2011, Thoroddsen2012, Tran2013}. The air that surrounds the falling object is squeezed out between the solid and pool surface during the final stages of impact, resulting in a local pressure build-up in the gas. This pressure will induce a small deformation of the liquid surface (Figure \ref{Setting}b), which will finally result in the entrapment of a tiny air bubble by the rupture of the enclosed air film (Figure \ref{Setting}c). For many applications these air bubbles are undesirable, and hence, the prediction of their sizes is of great importance.

There are mainly two types of theoretical approaches to determine the deformations of the liquid surface and predict of the size of the entrapped air bubble, namely full numerical solutions of the problem and scaling arguments \citep{Wilson1991, Hicks2011, Hicks2012, Mandre2012, Bouwhuis2012, Klaseboer2014}. Combined with experiments, these have lead to the observation that the size of an air bubble for impact of a liquid drop on a flat solid surface is determined by either the inertia of the liquid, or by the surface tension~\citep{Bouwhuis2012}. For increasingly high impact velocities, liquid inertia dominates and reduces the size of the air film at impact (`inertial regime'), while surface tension dominates for lower velocities or smaller drop sizes (`capillary regime'). The case of a solid sphere impacting on a pool leads to similar bubble entrapment, and, moreover, in the inertial regime the same scaling law (including the multiplicative prefactor) as for the impact of a drop on a solid was observed \citep{Marston2011,Tran2013}. Here, the final center height difference between the two surfaces, which is called the dimple height $H_d$, and the entrapped bubble volume $V_b$, respectively scale as

\begin{equation}
H_d \sim R~\St^{-2/3},\qquad V_b \sim R^3~\St^{-4/3}.
\label{scaling_inertial}
\end{equation}
Here $\St$ is the Stokes number, $\St=\rho_lUR/\eta_g$, in which $\rho_l$ is the density of the liquid, $R$ is the radius of the drop, $U$ is the impact velocity, and $\eta_g$ is the dynamic viscosity of the air. This scaling has been confirmed experimentally and numerically \citep{Marston2011, Hicks2011, Hicks2012, Mandre2012, Bouwhuis2012, Tran2013}. On the other hand, in the capillary regime (small velocities or small drops), the scaling analysis predicts \citep{Yiantsios1990,Bouwhuis2012}

\begin{equation}
H_d \sim R~\Ca^{1/2},\qquad V_b \sim R^3~\Ca,
\label{scaling_capillary}
\end{equation}
where $\Ca = \eta_gU/\gamma$ is the capillary number based on the gas properties and surface tension $\gamma$. 
The crossover in between the two regimes, at which the size of the entrapped air bubble is maximal, is found by equating the predictions for $H_d$ from Eqs. (\ref{scaling_inertial}) and (\ref{scaling_capillary}). Then, one finds $U_0\sim\eta_g^{1/7}\gamma^{3/7}/\left(\rho_l^{4/7}R^{4/7}\right)$, where $U_0$ is the crossover impact velocity, leading to maximal bubble entrapment. For an impacting water drop having a radius of $1~\mathrm{mm}$, this gives $0.07~\mathrm{m/s}$. Indeed, this is of the same order of magnitude as was observed experimentally, where the maximum bubble size was found around $0.25~\mathrm{m/s}$ (for ethanol drops) \citep{Bouwhuis2012}. Generically, for drops or spheres falling at their terminal velocity of a few meters per second, the impact thus takes place in the inertial regime, where the effect of surface tension can thus be neglected when focusing on the air bubble entrapment. Note that surface tension will enter during the rupture of the air film, which, however, appears to be on a different time scale. In \citet{Bouwhuis2012} it was experimentally found that, in the inertial regime, the bubble volume was fixed before the rupture of the air film.

In this paper, we analytically compute the initial deformations due to sphere impact onto a liquid pool in the \emph{inertial} regime, where the deflection of the liquid is limited by its inertia rather than by its surface tension. In experiments, there is generally not enough resolution to accurately detect these initial deformations, and therefore we use numerical simulations, bridging also towards larger deformations. By restricting ourselves to small deformations of the pool surface, we obtain detailed spatio-temperal information of the deflection as well as the dependence on experimental parameters. This provides a natural bridge between scaling theory, which lacks detailed information on the structure of the interface deflection, and profiles obtained by direct numerical simulations. Similar calculations were previously performed by \citet{Yiantsios1990} in the \emph{capillary} regime, recovering the scaling (\ref{scaling_capillary}). Hence, such a small-deformation theory gives an analytical foundation to the scaling laws, as well as detailed predictions for the shape of the deformation. Although the problem of a cushioning air layer has been solved by \citet{Wilson1991} for an `inertial' air layer, a similar insightful similarity analysis for the inertial (liquid) regime was not yet attempted.

The paper is organized as follows. Sec. \ref{sec:Formulation} starts with a dimensional analysis of the problem and shows the limiting cases when the gas can be described as a potential flow or as a viscous lubrication flow. This section also outlines the formalism based on which the interface deformations are computed. In Sec. \ref{sec:Results} we present the results for both viscous gas flow and potential gas flow. The analytical results are illustrated for a representative case of impact on a pool of water, with a sphere of radius $R=1~\mathrm{mm}$ and velocity $U=5~\mathrm{m/s}$, surrounded by air, as typical in experiments (inertial regime). Here we also provide a detailed comparison of our results with numerical simulations based on the Boundary Integral (BI) method, to validate our analysis and to investigate when the results start to deviate from the small-deformation regime. In Sec. \ref{sec:Conclusion} we conclude on the results in terms of air bubble entrapment.

\section{Formulation} \label{sec:Formulation}

The geometry of the problem is sketched in Figure \ref{Setting}: we consider a solid sphere (radius $R$) moving downwards towards the pool with a velocity $U$ (Figure \ref{Setting}a). The velocity of the sphere during its fall is assumed to be constant, i.e. we neglect the acceleration of gravity and the possible deceleration due to the gas flow. The movement of the air induces an increase of the gas pressure at the bottom of the sphere, which will then deflect the pool surface by a distance $\delta(r,t)$ (Figure \ref{Setting}b). The deformation $\delta$ is defined positive when the pool deflects downwards. For as long as the interface deflection is small with respect to the height of the gap, i.e. $|\delta|\ll h$, the problem can be solved by a perturbation analysis. In this section we first address the problem by dimensional analysis, and then provide the linearized formalism that allows computing the spatio-temporal evolution of the deflection $\delta(r,t)$. 


\subsection{Dimensional analysis} \label{subsec:Dimensional analysis}

Let us first consider the gas flow induced by the motion of the sphere. In the regime where the height of the gap is much larger than $R$, the sphere does not experience any influence of the pool. In that case, the Reynolds number of the gas flow is $\Re_g = \rho_gUR/\eta_g$, where $\rho_g$ is the density of air ($1.204~\mathrm{kg/m^3}$). 
However, as soon as the gap height becomes small, $h_0/R\ll1$, the air flow will be oriented mainly in the radial direction. As is typical for lubrication flows \citep{Reynolds1886}, one then has  to consider a different Reynolds number that is obtained from the radial component of the Navier Stokes equation. In terms of scaling laws this gives $\rho_gu_r^2/L\sim\eta_gu_r/h_0^2$, where $u_r$ is the typical radial gas flow velocity, and $L=\sqrt{Rh_0}$ is the length scale in the radial direction \citep{Hicks2012, Mandre2012,Bouwhuis2012,Klaseboer2014}. Applying mass conservation on the air gives $UL\sim u_rh_0$, and after elimination of $u_r$ one thus finds the relevant Reynolds number $\Re_{g,~lubr.}=\rho_gUh_0/\eta_g$. In the thin-gap regime, the relative influence of the viscosity and the inertia of the gas thus involves the gap thickness $h_0$ instead of the sphere radius $R$.  

It is instructive to evaluate these parameters for typical experimental values, such as spheres falling in air ($\rho_g=1.204~\mathrm{kg/m^3}$, $\eta_g=1.82\times10^{-5}~\mathrm{Pa~s}$) with $R=1~\mathrm{mm}$ and $U=5~\mathrm{m/s}$. The crossover from inertial to viscous gas flow, $\Re_{g,~lubr.}\sim 1$, arises when $h_0\sim 3~\mathrm{\mu m}$. This implies that there exists an ``inertial thin-gap regime", where $h_0/R < 1$ and $\Re_{g,~lubr.}>1$ at the same time. Only for the final stages of the impact, $h_0<3~\mathrm{\mu m}$, the gas can be described by a purely viscous flow. In the remainder, we therefore consider a potential flow analysis during two parts of the trajectory: the large-gap stage $h_0/R\gg1$, and the thin-gap stage $h_0/R\ll1$. The viscous flow is treated only in the final stages of impact, for which $h_0/R \ll 1$ and it is thus justified to reside to lubrication theory. The various limits will be worked out separately in Sec. \ref{sec:Results}.

The liquid pool is assumed to be a low-viscosity liquid and is treated for small amplitude deformations. These are essentially the same assumptions as for the propagation of linear surface waves \citep{Lamb1957}, where the amplitude is small with respect to the length scales of the problem. We focus on the ``inertial regime'' of impact, where the deformation is limited by the acceleration of the liquid and not by the surface tension of the liquid-air interface. Also the influence of gravity will be neglected in the theory; The Froude number based on the impact parameters $\Fr=U^2/\left(gR\right)$ is much larger than 1.

\subsection{From gas pressure to interface deflection} \label{subsec:From gas pressure to interface deflection}

The first step of the analysis is to compute the response of the liquid on a gas pressure $P_g$ for the different limiting cases (viscous/inertial gas), as discussed above. Since we set out to compute the initial deformation, we can compute $P_g$ assuming the liquid pool is undeformed -- the influence of a finite deflection is a correction at higher order in $\delta/h$. We assume axisymmetry and solve the equations in cylindrical coordinates $(r,z)$ (see Figure \ref{Setting}). The gas pressure will provide the boundary condition at the liquid pool, generating a liquid flow as described by the linearized Euler equation:

\begin{equation}
\frac{\partial {\vec{v}}}{\partial t} = -\frac{1}{\rho_l}\vec{\nabla}P_l,
\label{Euler}
\end{equation}
where ${\vec{v}(r,z,t)}$ is the velocity field in the liquid, and $P_l(r,z,t)$ is the pressure inside the liquid. The advection terms in the Euler equation are quadratic in velocity and therefore of higher order in $\delta/h$, in analogy to the wave analysis \citep{Lamb1957}. Without surface tension, the gas pressure provides the boundary condition for the liquid pressure

\begin{equation}
P_l(r,z\!=\!-\delta,t)\simeq P_l(r,z\!=\!0,t) = P_g(r,t),
\end{equation}
with the first equality again due to taking into account only leading order terms in $\delta/h$. The resulting deflection is given by the kinematic boundary condition:

\begin{equation}
\frac{\partial \delta}{\partial t}=-v_z|_{z=-\delta}-v_r|_{z=-\delta}\frac{\partial \delta}{\partial r} \simeq-v_z|_{z=0},
\label{kinematicBC}
\end{equation}
where $v_z|_{z=0}$ is the vertical velocity at the pool surface (to the lowest order in $\delta/h$). Substituting condition (\ref{kinematicBC}) into the vertical component of Eq. (\ref{Euler}) gives

\begin{equation}
\frac{\partial^2\delta}{\partial t^2} = \frac{1}{\rho_l}\left. \frac{\partial P_l}{\partial z}\right|_{z=0}.
\label{Euler2}
\end{equation}

The above equation shows that in order to compute $\delta(r,t)$, one requires a spatial derivative $\partial P_l/\partial z$. Hence, we need to find the pressure distribution inside the liquid that is induced by $P_g$ at the free surface. For an incompressible liquid this can be achieved by taking the divergence of Eq. (\ref{Euler}), which owing to $\vec{\nabla} \cdot \vec{v}=0$ reduces to $\nabla^2P_l=0$. As the boundary condition is axisymmetric, it is natural to express the pressure as the axisymmetric solution of the Laplace equation:

\begin{equation}
P_l(r,z,t)=\int\limits_0^\infty \widehat{P_g}(k,t) \JB(kr)\e^{kz}k\dx k,
\end{equation}
where the integration variable $k$ is the wave number, and $\JB(kr)$ is the Bessel function of the first kind with order $\nu=0$. The amplitude of the `modes' $\JB(kr)\e^{kz}$ is given by the Hankel transform of order 0 of the gas pressure $P_g(r,t)$,

\begin{equation}
\widehat{P_g}(k,t)=\int\limits_0^\infty P_g(r,t) \JB(kr)r\dx r.
\label{Hankel}
\end{equation}
Substituting this expression for the pressure into Eq. (\ref{Euler2}) gives

\begin{equation}
\frac{\partial^2\delta}{\partial t^2}(r,t) = \int\limits_0^\infty\frac{\widehat{P_g}(k,t)}{\rho_l}\JB(kr)k^2\dx k,
\label{invHankel}
\end{equation}
where we note an additional factor $k$ coming from the derivative of $\partial P_l/\partial z$.

The basic procedure for determining $\partial^2 \delta/\partial t^2$ from the gas pressure is now clear: one needs to find the Hankel transform of the gas pressure (Eq. (\ref{Hankel})), subsequently take the derivative of the result in the $z$-direction and evaluate the expression at $z=0$, and finally take the inverse Hankel transform (Eq. (\ref{invHankel})). In the following section we will perform these steps for the gas pressures computed in the limits of Stokes gas flow and inviscid gas flow.

\section{Results} \label{sec:Results}

\subsection{Stokes gas flow} \label{subsec:Stokes gas flow}

We now turn to the Stokes flow in the lubrication limit, which is valid for $\Re_{g,~lubr.}\ll1$ and $h_0/R\ll1$. In case of vanishing interface deformation, the gas pressure building up below an impacting sphere becomes \citep{Davis1986,Yiantsios1990}

\begin{equation}
P_g(r,t)=\frac{3\eta_gUR}{h_0^2\left(1+\frac{r^2}{2Rh_0}\right)^2} = \frac{3\eta_gU}{R}\left(\frac{R}{L}\right)^4F_1(u).
\label{PressureStokes}
\end{equation}
Here we factorized the result in dimensional parameters determining the magnitude of the pressure and a dimensionless function $F_1(u)$ that contains the spatial information of the pressure profile. For this, we introduced $L(t)=\sqrt{Rh_0(t)}$ as the relevant radial length scale, while the geometrical function reads

\begin{equation}
F_1(u) = \frac{1}{\left(1+\frac{1}{2}u^2\right)^2}; \qquad u(t) = \frac{r}{L(t)}.
\end{equation}
Note that in the limit of vanishing thickness $h_0$, the pressure tends to diverge, $P_g\sim h_0^{-2}$, while the width of the peak becomes increasingly small, $L\sim h_0^{1/2}$. These singular tendencies are regularized when the deformations of the surface become comparable to $h_0$, but yet, set the characteristic scales for the enclosed bubble volume.

We continue the analysis by inserting the gas pressure profile in Eq. (\ref{invHankel}), and find a closed form expression:

\begin{equation}
\frac{\partial^2\delta}{\partial t^2}(r,t) = \frac{3\eta_gU}{\rho_lRL}\left(\frac{R}{L}\right)^4G_1(u).
\label{Derivprofile_Stokes}
\end{equation}
Once more we recognize a dimensional prefactor that determines the scale of the acceleration, while the time-dependence follows from $L(t)$ and $u(t)$, and the spatial dependence through $G_1(u)$. The additional factor $1/L$ appearing in (\ref{Derivprofile_Stokes}) originates from the scaling $u=r/L$. The spatial similarity profile is $G_1(u)=\int\limits_0^\infty\widehat{F_1}\JB(ku)k^2\dx k$, where $\widehat{F_1}(k)$ is the Hankel transform of $F_1(u)$. The analytical expression for $\widehat{F_1}(k)$ is found to be

\begin{eqnarray}
\widehat{F_1}(k) = \sqrt{2}k\mathrm{K_1}(\sqrt{2}k),
\end{eqnarray}
where $\mathrm{K_1}(k)$ is the modified Bessel function of the second kind with order $\nu=1$, and the analytical expression for $G_1(u)$ is

\begin{equation}
G_1(u) = \frac{-8\mathrm{K}\left(\frac{u}{\sqrt{u^2+2}}\right)-\mathrm{E}\left(\frac{u}{\sqrt{u^2+2}}\right)+14\mathrm{E}\left(\frac{u}{\sqrt{u^2+2}}\right)}{\left(u^2+2\right)^{5/2}}.
\label{G_1}
\end{equation}
K and E are the complete elliptic integrals of the first and second kind, respectively.

\begin{figure}
\centering
\includegraphics[width=0.35\textwidth]{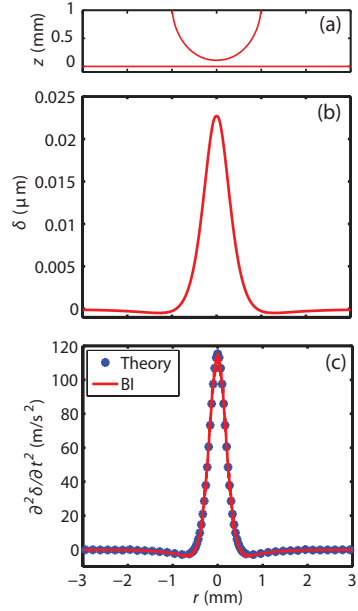}
\caption{Deflection of the pool interface for Stokes gas flow; $R=1~\mathrm{mm}$, $U=5~\mathrm{m/s}$, starting height of the (bottom of the) sphere $h_s=0.5~\mathrm{mm}$, current height: $h_0=0.1~\mathrm{mm}$. (a) Global view of the sphere and pool contours, (b) the pool deflection $\delta$ as a function of $r$, and (c) $\partial^2\delta/\partial t^2$ as a function of $r$. The solid red lines result from the Boundary Integral (BI) simulation. The theoretical result from Eq. (\ref{Derivprofile_Stokes}) has been superimposed in panel c (blue dots). Note the difference in scales on the vertical axes of panel a and b. The BI results agree perfectly with the theoretical predictions, as long as $|\delta|\ll h$.}
\label{Stokes_profiles_h0being100micron}
\end{figure}

\begin{figure}
\centering
\includegraphics[width=0.50\textwidth]{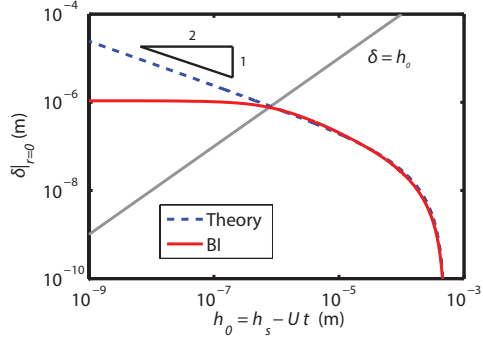}
\caption{Deflection of the pool interface on the axis, $\delta_{r=0}$, plotted against $h_0(t)$, for Stokes gas flow; $R=1~\mathrm{mm}$, $U=5~\mathrm{m/s}$, starting height $h_s=0.5~\mathrm{mm}$. The solid red line is the result from the BI simulation. The theoretical result from Eq. (\ref{diffeqStokes}) has been superimposed. After a start-up regime for large $h_0$, the deflection $\delta|_{r=0}$ converges towards a -1/2 power-law. The BI results perfectly agree with the theoretical predictions, until $\delta$ and $h_0$ become of comparable magnitude, pointed out by the crossing with the solid gray line $\delta|_{r=0}=h_0$. At that moment $\delta_{r=0}$ saturates to a constant value, which is the `dimple height' $H_d$ of \citet{Bouwhuis2012}.}
\label{Stokes_deltavsh0}
\end{figure}

To illustrate and validate our analysis, we compare the predicted profiles with the results obtained by Boundary Integral (BI) simulations \citep{Pozrikidis1997, Oguz1993, Bergmann2009}. The simulation method is the same as in \citet{Bouwhuis2012, Bouwhuis2013}: the liquid within the pool is described as a potential flow, while the pressure along the pool surface is explicitly calculated from the viscous lubrication equation for the gas flow. To be able to confirm our theoretical predictions in the inertial regime without the influences of surface tension and hydrostatics (which are both very small, as mentioned in the Introduction), $\gamma$ and $g$ are equal to zero in our simulations. In the limit of small deflection, the simulations should thus recover Eq. (\ref{Derivprofile_Stokes}).

Figure \ref{Stokes_profiles_h0being100micron}a shows the configuration on the scale of the sphere, for typical impact parameters for a sphere in air ($R=1~\mathrm{mm}$, $U=5~\mathrm{m/s}$). The interface deflection $\delta$ is shown in Figure~\ref{Stokes_profiles_h0being100micron}b, at the moment when the sphere is at a height $h_0=100~\mathrm{\mu m}$. At this time, $\delta\ll h_0\ll R$, for which we expect agreement between the BI results and our prediction from Eq. (\ref{Derivprofile_Stokes}). Figure \ref{Stokes_profiles_h0being100micron}c shows the acceleration $\partial^2\delta/\partial t^2$ versus $r$. The solid line is the result from the BI simulations and indeed gives perfect agreement with the prediction, represented by the dots. 

The actual deflection profile $\delta(r,t)$ can not be integrated explicitly from (\ref{Derivprofile_Stokes}), due to the time-dependence through $L$ and $u$. However, we can derive $\delta|_{r=0}$, the deflection of the pool surface on the axis, which does not involve $L(t)$. Using that $\partial/\partial t = -U\partial/\partial h_0$, we find

\begin{equation}
\frac{\partial^2\delta|_{r=0}}{\partial h_0^2}=\frac{3\eta_gG_1(0)}{\rho_lUR^2}\left(\frac{R}{L}\right)^5 = \frac{3\eta_gG_1(0)}{\rho_lUR^2}\left(\frac{R}{h_0}\right)^{5/2},
\label{diffeqStokes}
\end{equation}
where (\ref{G_1}) implies $G_1(0)=\frac{3}{8}\sqrt{2}\pi$. The solution of Eq. (\ref{diffeqStokes}) for $\delta|_{r=0}$ is subject to start-up effects as long as $h_0\sim h_s$, where $h_s$ is the initial height of the gap. If we let the initial height $h_s\rightarrow\infty$, we find 

\begin{equation}\label{eq:scaling}
\delta|_{r=0} \simeq {\rm \frac{3}{2}\sqrt{2}\pi} \frac{\eta_g}{\rho_l U} \left( \frac{R}{h_0} \right)^{1/2}.
\end{equation}
This predicts that the central height increases dramatically when $h_0$ decreases, as $\delta \sim h_0^{-1/2}$. Figure \ref{Stokes_deltavsh0} shows the BI result for $\delta|_{r=0}$ against $h_0$ (solid line), superimposed with the theoretical predictions (dashed line, taking into account the finite initial height $h_s$). Indeed, as soon as $h_0\ll h_s$, $\delta|_{r=0}$ converges to a $-1/2$ power law. As expected, the simulation results depart from the analytical prediction when $\delta \sim h_0$ (indicated by the solid gray line) and the lubrication approximation ceases to be valid. At this point, the deflection converges to a constant, which will be the final dimple height $H_d$. As stated in the Introduction, this will determine the dimple volume, and thus the entrapped air bubble volume, independently of the air film rupture process.

The current analysis provides a rigorous foundation for the scaling results obtained previously in \citet{Marston2011, Hicks2011, Hicks2012, Mandre2012, Bouwhuis2012}. There, the `dimple height' $H_d$ was observed to approach a constant value during the final stages of the impact. Figure \ref{Stokes_deltavsh0} shows that this height can be estimated from $\delta_{r=0}\sim h_0\sim H_d$. Using (\ref{eq:scaling}), this immediately gives

\begin{equation}\label{eq:stokesnumber}
H_d\sim\frac{\eta_g R^{1/2}}{\rho_lU H_d^{1/2}} \sim R~\St^{-2/3},
\end{equation}
where $\St=\rho_lUR/\eta_g$ is the Stokes number. The corresponding volume of the entrapped bubble volume then scales as

\begin{equation}\label{eq:stokesnumber_volume}
V_b \sim L^2H_d \sim RH_d^2 \sim R^3\St^{-4/3},
\end{equation}
where we use the common estimate that $L$ sets the lateral scale of the bubble. These are precisely the scaling predictions for the inertial regime (for Stokes gas flow), where the assumptions $H_d \sim \delta$ and $L \sim \left(H_dR\right)^{1/2}$ were further validated \citep{Marston2011, Hicks2011, Hicks2012, Mandre2012, Bouwhuis2012, Tran2013}.

%

\subsection{Potential gas flow}

As motivated in Sec. \ref{subsec:Dimensional analysis}, the inertial phase of the impacting sphere consists of two distinct stages: the large-gap regime $h_0\gg R$ and the thin-gap regime $h_0\ll R$. Below we separately treat both limiting cases analytically. We furthermore perform a numerical potential flow calculation for the full range of $h_0/R$, to validate the analysis and to show how the two stages are connected.

\subsubsection{Large-gap regime: $h_0\gg R$}

When the sphere is very far from the pool surface, the flow field can be described by the well-known potential flow field around a moving sphere of radius $R$. The introduction of the (undeformed) pool surface, however, requires that the gas velocity has no vertical component, or $v_z|_{z=0}=0$. This boundary condition can be satisfied using the `method of images', corresponding to two approaching spheres having radius $R$ with approaching velocity $U$ towards a mirroring horizontal line ($z=0$). Applying the superposition of the potentials for the two moving spheres, one obtains the potential

\begin{equation}
\phi(r,z,t) = \frac{UR^3}{2}\left[\frac{\left(z-R-h_0\right)}{\left(r^2+\left(z-R-h_0\right)^2\right)^{3/2}}-\frac{z+R+h_0}{\left(r^2+\left(z+R+h_0\right)^2\right)^{3/2}}\right].
\label{PotentialPotential}
\end{equation}
It is important to realize that the introduction of the second moving sphere not only influences the flow around $z=0$, but also gives a small, unwanted velocity on the boundary of the original sphere. In the limit of very large gaps, $R/h_0 \ll 1$, this correction becomes negligible and (\ref{PotentialPotential}) gives the asymptotically correct potential.

We now extract the gas pressure profile on the level of the pool surface $z=0$, by applying the unsteady Bernoulli equation:

\begin{equation}
P_g(r,t) = \rho_gU^2\left[\left(\frac{R}{\zeta}\right)^3 F_2(u) + \frac{9}{2}\left(\frac{R}{\zeta}\right)^6 F_3(u)\right] \simeq \rho_gU^2\left(\frac{R}{\zeta}\right)^3F_2(u).
\label{PotentialPressure}
\end{equation}
Here, $\zeta(t)=R+h_0(t)=R+h_s-Ut$, the radial direction is scaled as $u(t)=r/\zeta$, while the spatial profiles are

\begin{eqnarray}
F_2(u) &=& \frac{2-u^2}{\left(1+u^2\right)^{5/2}}; \\
F_3(u) &=& \frac{-u^2}{\left(1+u^2\right)^5}.
\end{eqnarray}
Since (\ref{PotentialPotential},\ref{PotentialPressure}) are only valid for $h_0\gg R$, we only keep the dominant first term in (\ref{PotentialPressure}). Note that the width of the pressure peak is now set by the scale $\zeta=h_0+R$. This can be contrasted with the width in the thin-gap limit, $L=\sqrt{Rh_0}$, which becomes very narrow.

Next, from (\ref{PotentialPressure}) we can compute the induced acceleration profile using (\ref{invHankel}):

\begin{equation}
\frac{\partial^2\delta}{\partial t^2}(r,t) = \frac{\rho_gU^2}{\rho_l\zeta}\left(\frac{R}{\zeta}\right)^3G_2(u).
\label{Derivprofile_potential}
\end{equation}
One recognizes a dimensional prefactor that is separated from the spatio-temporal dependence. The function $G_2(u) = \int\limits_0^\infty \widehat{F_2}\JB (ku)k^2\dx k$ is the spatial similarity profile, where $\widehat{F_2}(k)$ is the Hankel-transform of $F_2(u)$. For $G_2(u)$ we did not find any analytical expression, but one can numerically calculate the given integral (cf. Figure \ref{potential_profiles_h0being10mm}).

Once again, we can analytically compute the behavior of the central deflection, $\delta|_{r=0}$:

\begin{equation}
\frac{\partial^2\delta|_{r=0}}{\partial h_0^2}=\frac{\rho_gG_2(0)}{\rho_lR}\left(\frac{R}{\zeta}\right)^4.
\label{diffeqpotential}
\end{equation}
Recalling that $\partial/\partial h_0=\partial/\partial \zeta$ and $\zeta\rightarrow2R$ for $h_0\rightarrow R$, this implies that the final $\delta_{r=0}$ scales as $\rho_gR/\rho_l$. In contrast to the result for viscous flow, the typical deformation versus $h_0$ depends only on the density ratio $\rho_g/\rho_l$, but not on the impact velocity. While the density ratio is typically small, we anticipate that the resulting deflection for a millimeter-sized sphere can be a few microns. This is actually comparable to typical deflections in the viscous lubrication phase. However, the pool is not deformed locally over a small width $\sqrt{Rh_0}$, but over the scale of the entire sphere, and therefore it will be of little consequence for the formation of the dimple and the size of the entrapped air bubble.

%

%
%

\subsubsection{Thin-gap regime: $h_0\ll R$}

In the inertial thin-gap limit, the gas is squeezed out mainly in the radial direction. To predict the pressure profile for this stage of the impact, we use the depth-integrated continuity equation \citep{Snoeijer2009,Bouwhuis2013}

\begin{equation}\label{eq:depth}
\frac{\partial h}{\partial t}+\frac{1}{r}\frac{\partial}{\partial r}\left(rh\overline{u}_r\right)=0,
\end{equation}
where $\overline{u}_r(r,t)$ is the height-averaged radial gas velocity in the gap. Assuming a plug flow that does not depend on the $z$-coordinate, this average simply gives $\overline{u}_r(r,t) = u_r(r,t)$. This analytical description is similar to what has been done by \citet{Wilson1991}, who also studied cushioning air-layers at solid-liquid impact in the inertial thin-gap regime, though in 2D Cartesian coordinates, for general shapes of the impacting solid. In the present case, the bottom of the impacting solid sphere can be described as $h=h_0(t)+r^2/\left(2R\right)$, and thus, $\partial h/\partial t=\partial h_0/\partial t=-U$. Hence, we can integrate (\ref{eq:depth}) to find

\begin{equation}
u_r=\overline{u}_r=\frac{Ur}{2h_0\left(1+\frac{r^2}{2Rh_0}\right)}.
\label{u_r}
\end{equation}
The velocity profile (\ref{u_r}) has a local maximum at $r=\sqrt{2Rh_0}$, and vanishes for $r=0$ and $r=\infty$. Substituting the profile into the radial component of the Euler equation and integrating over $r$ gives the gas pressure:

\begin{equation}
P_g(r,t) = \frac{\rho_gU^2R}{2h_0}\left(\frac{1+\frac{r^2}{4Rh_0}}{\left(1+\frac{r^2}{2Rh_0}\right)^2}\right) = \frac{\rho_gU^2}{2}\left(\frac{R}{L}\right)^2F_4(u),
\label{PotentialPressure_thingap}
\end{equation}
with $L(t)=\sqrt{Rh_0(t)}$, $u(t)=r/L$, and

\begin{equation}
F_4(u)=\frac{1+\frac{1}{4}u^2}{\left(1+\frac{1}{2}u^2\right)^2}.
\end{equation}
Note that the geometry of the thin-gap again gives rise to a highly localized pressure profile of a width $\sqrt{Rh_0}$. The gas pressure again tends to diverge as $h_0 \rightarrow 0$, but more slowly than in the viscous case: the inertial gas pressure in the thin-gap-limit is proportional to $1/h_0$, in contrast to the more singular scaling for the viscous gas flow scenario, $1/h_0^2$.

From (\ref{invHankel}) we deduce the pool surface acceleration

\begin{equation}
\frac{\partial^2\delta}{\partial t^2}(r,t) = \frac{\rho_gU^2}{2\rho_lL}\left(\frac{R}{L}\right)^2G_4(u),
\end{equation}
where $G_4(u) = \int\limits_0^\infty \widehat{F_4}\JB (ku)k^2\dx k$, with $\widehat{F_4}(k)$ the Hankel-transform of $F_4(u)$. At the origin $r=0$, this reduces to

\begin{equation}
\frac{\partial^2\delta|_{r=0}}{\partial h_0^2} = \frac{\rho_gG_4(0)}{2\rho_lR}\left(\frac{R}{L}\right)^3.
\end{equation}
Just like in case of the large-gap regime, the central deflection has no dependence on impact velocity. Solving gives $\delta_{r=0}\sim h_0^{1/2}+~\mathrm{integration~constants}$. From this we conclude that in the inertial thin-gap limit, the pressure tends to \emph{diverge} for $h_0\rightarrow0$, but the deflection $\delta$ \emph{converges}. Contrarily to the final stages in the case of viscous gas flow, the inertial gas pressure is not sufficiently singular to induce a strongly enhanced deflection. The integration constants depend on the full history of the impact process, which thus involves the dynamics during the preceding large-gap regime. To predict the actual deflection during the final stages of sphere impact, it is thus not sufficient to consider the large-gap or thin-gap regime of the potential gas flow problem, but requires numerical simulation of the full impact process over all $h_0/R$.

\subsubsection{Numerical simulations} \label{subsubsec:Numerical simulations}

Simulating the potential gas flow impact process using the BI technique calls for a different approach with respect to the case of Stokes gas flow. The reason is that we require the gas pressure over the full range of gap thickness, including $h_0\sim R$, for which no analytical solution for the gas pressure is available that can serve as a boundary condition for the liquid pool. As a consequence, the gas phase must be also computed numerically, which we achieve using the Boundary Integral code. We thus need to run two separate simulations. The process is started by a BI simulation of a solid sphere impacting towards an \emph{undeformed} surface, with in between a potential gas flow. From this simulation, the gas pressure profile along the pool surface ($z=0$) is extracted. In the second BI simulation, this pressure is applied on a deformable pool surface, from which we eventually determine the resulting pool deflections. This is again a valid method as long as $\delta/h\ll 1$. The pressure data is transmitted from the first simulation to the second simulation through an extensive data file. Note that by doing two separate simulations, one needs to take into account the different length scales during the impact process (for $h_0=10~\mathrm{mm}\rightarrow100~\mathrm{nm}$), implying very sensitive local node spacings and time dependencies. This was achieved by adapting the node spacing and time steps to ensure  convergence of the numerical results. 

\begin{figure}
\centering
\includegraphics[width=0.35\textwidth]{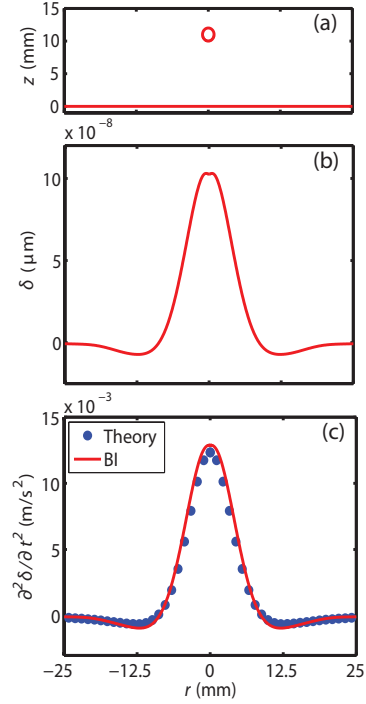}
\caption{Deflection of the pool interface for potential gas flow in the limit $h_0\gg R$; $R=1~\mathrm{mm}$, $U=5~\mathrm{m/s}$, $h_0=h_s=10~\mathrm{mm}$ (thus, $h_0/R=10$). (a) Global view plot of the sphere and pool contours, (b) $\delta$ against $r$, and (c) $\partial^2\delta/\partial t^2$ against $r$. The solid red lines result from the BI simulation. The theoretical result from Eq. (\ref{Derivprofile_potential}) has been superimposed in panel c (blue dots). Note the difference in scales on the vertical axes of panel a and b. The BI results are nicely agreeing with the theoretical predictions, until $h_0/R$ becomes of order 1.}
\label{potential_profiles_h0being10mm}
\end{figure}

\begin{figure}
\centering
\includegraphics[width=0.80\textwidth]{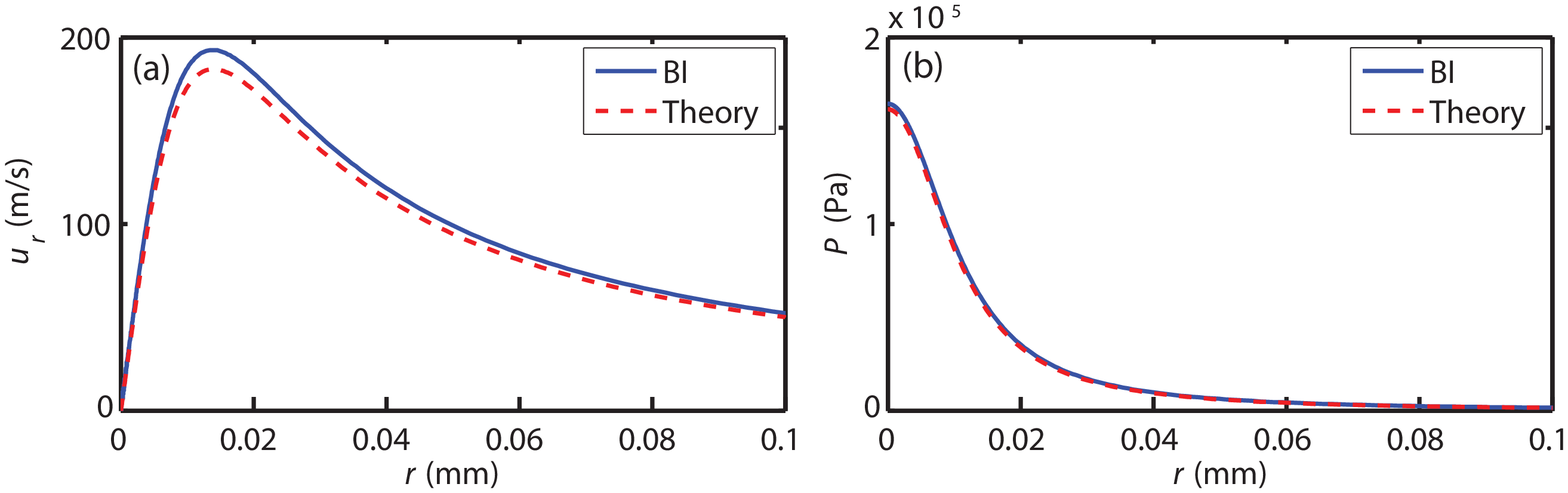}
\caption{Inertial gas flow in the thin-gap limit. Theoretical prediction (red dashed line) and BI gas flow simulation result (blue solid line) of the velocity (panel a) and pressure (panel b) profile within the gas; $R=1~\mathrm{mm}$, $U=5~\mathrm{m/s}$, $h_s=h_0=100~\mathrm{nm}$ (thus, $h_0/R=10^{-4}$). We find very good agreement between the theoretical predictions and the BI results.}
\label{potential_velprofileandpressure_h0being100nm}
\end{figure}

\begin{figure}
\centering
\includegraphics[width=0.50\textwidth]{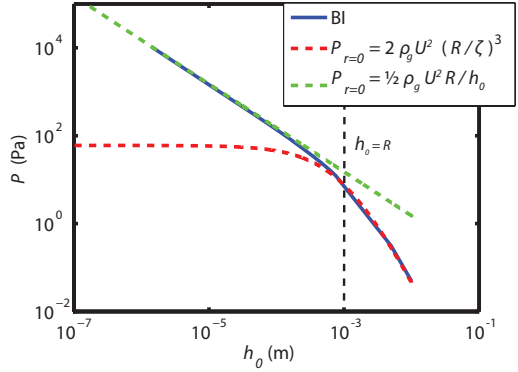}
\caption{Behavior of the gas pressure on the axis, $P_{r=0}$, plotted against the gap height $h_0(t)$, for potential gas flow. $R=1~\mathrm{mm}$, $U=5~\mathrm{m/s}$, $h_s=10~\mathrm{mm}$.  The red dashed line is the theoretical prediction in the regime $h_0\gg R$; The green dashed line is the theoretical prediction in the regime $h_0\ll R$. The dashed line points out the crossover $h_0=R$. The BI gas flow simulation result (blue solid line) indeed follows these predicted behaviors in the corresponding regimes, with a crossover at $h_0\sim R$.}
\label{potential_Pvsh0}
\end{figure}

\begin{figure}
\centering
\includegraphics[width=0.50\textwidth]{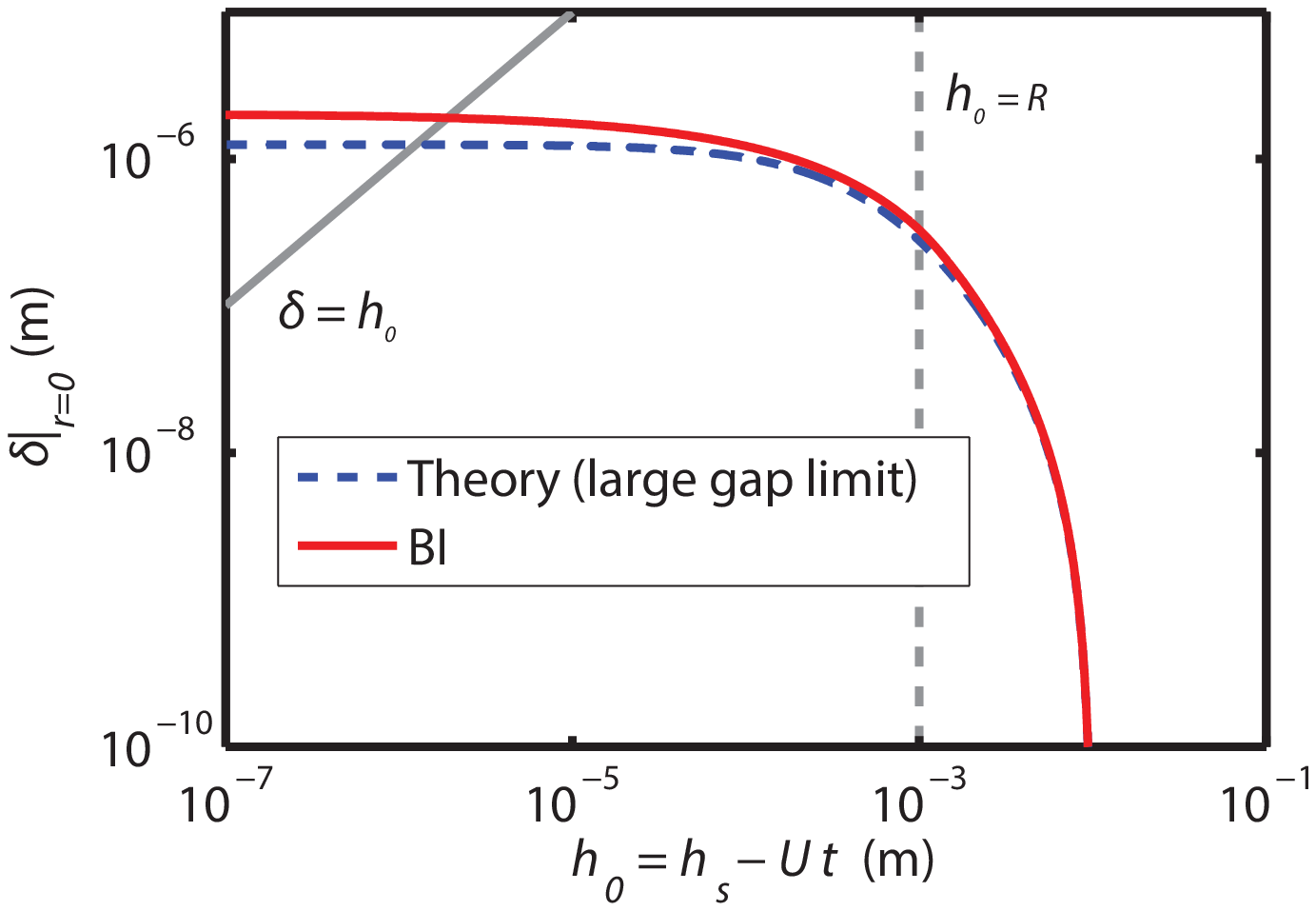}
\caption{Deflection of the pool interface on the axis, $\delta_{r=0}$, plotted against the gap height $h_0(t)$, for potential gas flow; $R=1~\mathrm{mm}$, $U=5~\mathrm{m/s}$, $h_s=10~\mathrm{mm}$. The solid red line is the result from the BI solid sphere on liquid pool simulations (Sec. \ref{subsubsec:Numerical simulations}). The theoretical result from Eq. (\ref{diffeqpotential}) for the large-gap regime has been superimposed (blue dashed line). $\delta|_{r=0}$ saturates to a constant. The BI results perfectly agree with the large-gap predictions in the regime $h_0\gg R$. In the regime $h_0\ll R$, $\delta_{r=0}$ deviates from this prediction, but the difference is relatively small. The dashed gray line points out the crossover $h_0=R$; The solid gray line points out $\delta|_{r=0}=h_0$.}
\label{potential_deltavsh0}
\end{figure}

Figure \ref{potential_profiles_h0being10mm}a and b respectively show the configuration on the length-scale of the sphere, and the interface deflection $\delta(r,t)$ for $R=1~\mathrm{mm}$, $U=5~\mathrm{m/s}$, and $h_0=h_s=10~\mathrm{mm}$ (i.e., the large-gap regime). Figure \ref{potential_profiles_h0being10mm}c shows the acceleration profile at the corresponding time, and is observed to agree very well with the asymptotic result of Eq. (\ref{Derivprofile_potential}) (blue dots). The very small difference between the BI result and the theoretical predictions can be explained by the fact that $h_s/R=10$, implying an expected difference of about $10\%$ between the theory and the numerical simulations. We remark that the corresponding deformation (Figure \ref{potential_profiles_h0being10mm}b) is very small, as we look to the very initial deformations in the start-up regime. The demand $h_0/R\gg 1$ implies a large initial gap height, which, for the parameter values chosen in Fig. \ref{potential_profiles_h0being10mm} to validate the asymptotics, corresponds to nonphysically small deflections. The sensitiveness of the very small pool deflection gave rise to switch off the normally used smoothing procedure within the simulations \citep{Oguz1993}, such that a tiny instability remained visible around the axis, $r=0$. We confirmed that this instability has a numerical origin and that it does not influence the result on the scale of the deformations. The thin-gap regime is analyzed in Figure \ref{potential_velprofileandpressure_h0being100nm}. We again find very good agreement between the analytical gas velocity profile (panel a) and the pressure profile (panel b) and the BI results (here $h_0=100~\mathrm{nm}$).

The crossover between the large-gap and thin-gap limits is illustrated in Figure \ref{potential_Pvsh0}, showing the gas pressure on the symmetry axis $r\!\!=\!\!0$. As predicted, in the limit $h_0/R\gg1$ the pressure calculated by BI (blue line) equals $2\rho_gU^2\left(R/\zeta\right)^3$ (red dashed line), and in the limit $h_0/R\ll1$ the pressure equals $\rho_gU^2R/(2h_0)$ (green dashed line). This confirms the validity of the analytical approaches. Finally, we investigate the deflection of the pool that is induced by the numerically obtained gas pressure. Figure \ref{potential_deltavsh0} shows the deflection at $r\!\!=\!\!0$, the inertial (gas) counterpart of Figure \ref{Stokes_deltavsh0}. As expected, $\delta_{r=0}$ deviates from the large-gap prediction in the small-gap regime, though the deviation is not very large. This means that, despite the fact that the gas pressure tends to diverge for $h_0\rightarrow0$, the influence of the inertial thin-gap limit remains relatively small. For this particular example, it enhances the deflection by less than a factor $2$. This is also one of the reasons that we do not show the corresponding theoretical profile for $\partial^2 \delta/\partial t^2$, which in principle could again be directly calculated from the pressure profile \footnote{A second reason is that, in the numerical simulations, the very small gap height of $100~\mathrm{nm}$ needs a very high local node density on both the pool surface and the sphere surface; The difference in length scales of $R$ and $h_0$ is four decades, which is very challenging. This induces very small time steps to be able to calculate a fair second derivative of the deflection profile in time. In addition, the pressure along the pool surface needs to be extracted from a prior solid-sphere-on-solid-surface simulation (through an extensive data file), which makes the discretization more complicated.}.

The large-gap prediction for the final $\delta_{r=0}$ is thus satisfactory, and we conclude with the following scaling law for the resulting dimple height $H_d$ for the inertial gas scenario as was concluded from Eq. (\ref{diffeqpotential}):

\begin{equation}
H_d\sim R\frac{\rho_g}{\rho_l}.
\end{equation}
This dimple height is independent of the impact velocity of the sphere. Since the surface deformation is the sum of the deformations in the both the large-gap and the thin-gap limit, it is unclear what the correct radial and axial length scales are that lead to the volume of the pinched bubble.

\section{Conclusion} \label{sec:Conclusion}

We performed a perturbation analysis to investigate the initial deflections of a liquid surface, induced by the approach of an impacting solid sphere. The analysis assumes the deflection is limited by the inertia of the liquid pool (i.e., not by its surface tension) and we consider two natural limits for the surrounding medium: Stokes gas flow and potential gas flow. We obtained a quantitative prediction for the pool surface deflection, which was validated numerically, and recovered previously proposed scaling laws for bubble entrapment. 

While the `cushioning' of an inertial gas layer had been analyzed before \citep{Wilson1991}, most recent work on liquid or solid impact assumes a viscous gas layer. Surprisingly, our analysis reveals that inertial and viscous cushioning both lead to a pool deflection of the order $1~\mathrm{\mu m}$, for typical experimental conditions. However, the Stokes gas pressure strongly tends to diverge for $h_0\rightarrow0$, much more strongly than during the inertial gas phase. In addition, this viscous lubrication pressure profile is very localized, while most of the inertial deflection is generated during the initial phase where the pool deflection is spread over the entire width of the sphere. This explains why the experimental results on bubble entrapment are in close agreement with the scaling law (\ref{eq:stokesnumber_volume}) \citep{Tran2013}, while in addition (\ref{eq:stokesnumber}) was validated for the case of a liquid drop impact at a solid \citep{Marston2011, Hicks2011,Hicks2012,Mandre2012,Bouwhuis2012,Tran2013}, which are all based on the viscous lubrication regime.

For completeness, we will summarize the possible scenarios for impact of a sphere onto a pool, that can be achieved for different experimental parameters. Assuming an initially high Reynolds number based on the size of the impacting object $R$, the dynamics will exhibit two different types of crossover: a geometric crossover based on the relative thickness of the gap $h/R$, and a crossover from inertial to viscous gas flow. The order in which these crossovers occur depends on the parameters of the problem. In our numerical examples we assumed one first reaches the thin-gap regime, before the lubrication Reynolds number (based on the gap thickness $h$), becomes smaller than unity. This order can be reversed for impact at smaller velocities or for a sphere sinking in a more viscous medium. In that case, however, one needs to bear in mind that the influence of the pool surface tension will become more important, corresponding to the capillary impact regime. In this case, the thin film potentially has time to drain out before a bubble is formed, making the entrapment process more complex \citep{Klaseboer2000,Yoon2005}.


In this work, we have elaborated on the impact of a solid sphere onto a liquid surface. Similar perturbation analysis can be performed for drop impact on a solid, or drop impact on a pool, though details will be different. This explains why the same scaling laws are observed in all these cases.

\section*{Acknowledgments}

We gratefully acknowledge Stephen Wilson and Hanneke Gelderblom for insightful discussion. This work was supported by STW and NWO
through a VIDI Grant No. 11304.

\bibliographystyle{jfm}
\bibliography{refs}

\end{document}